\documentclass[prl,twocolumn]{revtex4}
\usepackage{amsthm}
\usepackage{amsmath}
\usepackage{amssymb}
\usepackage{amsfonts}
\usepackage{booktabs}
\usepackage{bbm}
\usepackage{hyperref}

\def\={\ =\ }

\newcommand{\Z}{\mathbb{Z}}

\newcommand{\N}{\mathbb{N}}

\begin{document}

\title[\hfill\ \ ]{The Heisenberg XX spin chain and low-energy QCD}
\author{David P\'{e}rez-Garc\'{\i}a}
\author{Miguel Tierz}
\affiliation{Departamento de An\'{a}lisis Matem\'{a}tico and Instituto de Matem\'{a}tica
Interdisciplinar, Universidad Complutense de Madrid, 28040 Madrid, Spain}

\begin{abstract}
By using random matrix models we uncover a connection between the low energy sector of four dimensional QCD at finite volume and
the Heisenberg XX model in a 1d spin chain. This connection allows to relate crucial
properties of QCD with physically meaningful properties of the spin chain,
establishing a dictionary between both worlds. We predict for the spin chain a third-order phase transition and a Tracy-Widom law in
the transition region. 
We finally comment on possible numerical implications of the connection as well as on possible experimental implementations.
\end{abstract}

\maketitle

The strong interaction is the fundamental force of nature which describes
the interaction between quarks and gluons, the elementary constituents of
hadronic matter. It is described by quantum chromodynamics (QCD), a $SU(3)$
Yang-Mills theory with a number of distinctive properties such as asymptotic
freedom \cite{AF}, which correctly describes that the interaction between
particles becomes asymptotically weaker as distance decreases and energy
increases. This crucial property, in agreement with Bjorken scaling and
experimental data, is due to the negativity of the $\beta $ function
describing the variation of the coupling constant of the theory under the
renormalization group flow \cite{AF}. Phenomenology also tells us that the up and down quarks are very light.
The case of massless quarks implies some additional symmetries, named chiral
symmetries, which would allow separate transformations between the
left-handed quarks and the right-handed ones. Such behavior is not observed
and hence, in a realistic QCD, the chiral symmetry must be spontaneously
broken. Low energy QCD, which is the regime we are interested in, is deeply related to the notion of chiral
symmetry breaking and it can be explored
with chiral perturbation theory \cite{Wein79,GH84,GH87}. Recall that quarks interact weakly at high energies and strongly
at low energies and, therefore, the low-energy regime is described by
non-perturbative physics. Finally, chief among the features of QCD is the confinement of quarks into hadrons,
either mesons ($q\overline{q}$) or baryons ($qqq$). Confinement in a gauge
theory is usually probed by studying the behavior of Wilson loops observables \cite%
{Wilson}.

There has been a lot of theoretical and numerical approaches to analyze QCD and related gauge theories, such as effective field theory and chiral perturbation theory 
\cite{Wein79,GH84}, lattice gauge theory \cite{Creutz,Smit}, the light-cone
quantization \cite{Light-cone}, gauge-string duality and AdS/CFT approaches 
\cite{Kor}, etc. Very recently, and motivated by an idea of Feynman \cite{Feyn}, a new route
has appeared to understand Abelian and non-Abelian gauge theory: simulating
it in a different controllable quantum system, such as cold atoms in optical
lattices \cite{Lewrev,Lewbook}. Some first steps for different quantum field
theories (QFT) have been carried out in \cite%
{Preskill,Preskill2,Lew,Lew2,Zoller,Zoller2,ZCR1,ZCR2}.

\begin{table}
\begin{tabular}{| l | l |}
\toprule
\hline
\cmidrule(r){1-2}
{\bf Low energy QCD}    & {\bf Thermal correlations} \\
& {\bf (XX model)}\\
\midrule
\hline
      number of flavors      & number of particles       \\
\hline
topological sector  & ket vs. bra shift       \\
\hline
$\theta$ angle & projection onto momentum $\theta$\\
       \hline
 different matter content     & different boundary conditions    \\
 \hline
on the lattice      & longer range interactions      \\
\hline
\bottomrule
\end{tabular}
\caption{{\it Dictionary} relating properties of QCD with 
properties of the thermal averages in the XX spin chain.}
  \label{tab:dictionary}
\end{table}

In this paper, we initiate a different, but somehow related approach. By
combining a result of Leutwyler and Smilga \cite{LS} (based on the previous
seminal work on chiral perturbation theory \cite{GH84,GH87}) with a result
of Bogoliubov et al. \cite{Bog, Bog2, BC,B} we uncover a mapping between the
low energy sector of QCD with thermal correlation functions in the 1D
Heisenberg XX model or, via a Jordan-Wigner transformation \cite{LSM},
thermal correlation functions in a 1D free fermion system. The connection is
made by relating both objects to a random matrix model \cite{RM}. Building
upon this starting point, we are able then to relate crucial properties of
QCD with physically meaningful properties of the spin chain. For instance we
show that the number of flavors in QCD corresponds
to the number of particles (spins down) in the 1d chain; (2) the topological
charge in QCD is associated with the signature that topological 2D systems leave on their boundary theory \cite{Haldane,Qi,Schuch1,Schuch2}; (3) different matter content, such as
Majorana fermions, corresponds to different boundary conditions in the spin
chain; or that (4) putting QCD on the lattice enforces the addition of
next-to-nearest neighbor terms in the spin chain Hamiltonian.
The connection
allows us to uncover also a third order phase transition in the XX model
since, again via random matrix models, one can relate both low-energy QCD
and the thermal correlation functions of the XX chain with the so called
Gross-Witten model, a 2d Yang-Mills theory with gauge group $U(N)$ and no
matter fields which has a third order phase transition in the limit $%
N\rightarrow \infty $ \cite{GW}. 
Finally the connection opens the possibility of using  numerical methods coming from spin chains \cite{T-DMRG} to give good estimates for the partition function of low-energy QCD. It also opens a way to measure this partition function or to observe the Gross-Witten phase transition experimentally.

\subsection{Low energy QCD as a random matrix model}

Let us start by describing the derivation in \cite{LS}, which applies the
ideas of effective field theory \cite{Wein79} to the study of the meson
sector of QCD \cite{GH87,LS}. Recall that the main idea of an effective
field theory approach is to integrate out the heavy degrees of freedom
(the most massive fields) of the theory. This is implemented to study
the low-energy (meson) sector of QCD, through chiral perturbation theory ($%
\chi $PT) \cite{GH84} with the quark and gluon fields of QCD replaced by a
set of pion fields $U(x)$, which describe the degrees of freedom of the
pseudo Nambu-Goldstone bosons. The effective Lagrangian depends only on the
pion fields and its derivatives 
\begin{equation*}
\mathcal{L\rightarrow L}_{\mathrm{eff}}(U,\partial U,...)=\mathcal{L}_{%
\mathrm{eff}}^{\left( 0\right) }+\mathcal{L}_{\mathrm{eff}}^{\left( 2\right)
}+\mathcal{L}_{\mathrm{eff}}^{\left( 4\right) }+...
\end{equation*}%
Chiral symmetry provides a tight
constraint to the form of these terms and, in particular, the first term $%
\mathcal{L}_{\mathrm{eff}}^{\left( 0\right) }$ is just a constant which is
the vacuum energy of QCD in the chiral limit. The first non-trivial term is $%
\mathcal{L}_{\mathrm{eff}}^{\left( 2\right) }$ and is given by \cite%
{Wein79,GH84}%
\begin{equation}
\mathcal{L}_{\mathrm{eff}}^{\left( 2\right) }=\frac{1}{4}F^{2}\mathrm{tr}%
\left\{ \partial _{\mu }U^{\dag }\partial ^{\mu }U\right\} +\frac{1}{2}%
F^{2}\Sigma \mathrm{tr}\left\{ M\left( U+U^{\dag }\right) \right\}, \label{Lag}
\end{equation}%
where $M$ is the matrix that contains the masses of the quarks (quark mass matrix) and that will be taken below to be a multiple of the identity matrix, $F$ is the pion decay constant, and $%
\Sigma $ is the chiral condensate (which describes the spontaneous chiral
symmetry breaking). 

Chiral perturbation theory gives an expansion in powers of $M$, temperature $T$ and inverse length $1/L$ at fixed $\Lambda _{QCD}$ and at fixed ratios, $M/T^{2}$ and $LT$. The scales $T$ and $1/L$ are treated as small quantities
of order $p$, where $p$ is the momentum of the pions, whereas the quark mass matrix counts as a quantity of order $p^{2}$ \cite{GH87}. The most general gauge invariant expression consistent with the symmetries of QCD that can be formed within the effective theory at $\mathcal{O}(p^{2})$ is given by (\ref{Lag}). This Lagrangian holds under the condition for the volume $V^{1/4} \gg 1/\Lambda_{QCD}$ where $\Lambda _{QCD}$ is the length scale of QCD \cite{GH87}. In this way only the Goldstone modes contribute to the mass-dependence of the partition function \cite{GH87}.

In addition, for quark masses for which the Compton
wavelength of the Goldstone modes is much larger than the size of the box $1/m_{\pi }\gg V^{1/4}$. \footnote{$m_{\pi }$ denotes the mass of the lightest Goldstone boson and can be expressed in terms of the low-energy constants $F,$ $\Sigma $ and the masses of the quarks. See \cite{LS}
or the review \cite{V-rev} for details.}. This is known as the epsilon regime of QCD since it is an expansion in terms of $\varepsilon ^{2}\sim \frac{m_{\pi}}{\Lambda _{\mathrm{QCD}}}$ \cite{GH87}. The fluctuations of the zero momentum modes of the pion fields dominate the fluctuations 
of the nonzero momentum modes and only the former are taken into account in the thermal average \cite{LS}. 
These two conditions on the volume are also referred to as the kinetic domain \cite{V-rev} since the kinetic
term of the chiral Lagrangian can be ignored. The low-energy partition function is then \cite{GH87}
\begin{equation*}
Z_{\text{eff}}(M,\theta )=\int_{U\in SU(N_{f})}dU\exp \left( V\Sigma \;\mathrm{Re}\,%
\mathrm{tr}\{MU^{\dag }\}e^{i\theta /N_{f}}\right), 
\end{equation*}%
since only the constant fields contribute to its mass dependence.

Note that the inverse temperature $\beta$ of the gauge theory does not appear since, in the low-energy effective
field theory, one can absorb it in the low-energy constants \cite{GH87}. The appearance of the $\theta $ parameter is because, due to the explicit
breaking of the axial symmetry $U_{A}(1)$, one is naturally led to also
consider the addition of a theta term to the original QCD Lagrangian \cite{LS}%
\begin{equation*}
\mathcal{L}_{\mathrm{\theta}}\mathcal{=-}\frac{i\theta }{%
32\pi ^{2}}F_{\mu \nu }^{a}\widetilde{F}_{\mu \nu }^{a},
\end{equation*}%
\newline
where the field strength and its dual are given by \cite{gauge-book}
\begin{equation*}
F_{\mu \nu }^{a}=\partial _{\mu }A_{\nu }^{a}-\partial _{\nu }A_{\mu
}^{a}+f_{abc}A_{\mu }^{b}A_{\nu }^{c}\text{ \ \ \ \ \ \ \ \ \ }\widetilde{F}%
_{\mu \nu }^{a}=\frac{1}{2}\varepsilon _{\mu \nu \alpha \beta }F^{\alpha
\beta },
\end{equation*}%
with $f_{abc}$ the structure constants of the gauge group $SU(N_{c})$. This is the same
topological term that appears in topological insulators \cite{top}. The
topological charge%
\begin{equation*}
\nu =\frac{1}{32\pi ^{2}}\int F_{\mu \nu }^{a}\widetilde{F}_{\mu \nu
}^{a}d^{4}x
\end{equation*}%
characterized by the integer $\nu$, is a topological invariant. The low-energy QCD partition function is then written as 
\cite{LS} 
\begin{equation}\label{qcd-theta}
Z_{QCD}^{\mathrm{eff}}(M,\theta )=\sum_{\nu =-\infty }^{\infty }e^{i\nu \theta }Z_{\nu }^{\mathrm{eff}}(M).
\end{equation}%
The effective partition function at fixed $\nu $ follows then from Fourier
inversion \cite{LS}%
\begin{equation}
Z_{\nu ,N_{f}}^{\mathrm{eff}}(M)=\int_{U(N_{f})}dU\left( \det (U)\right)
^{\nu }\exp \left( \frac{V\Sigma }{2}\mathrm{Tr}\left[ \left( M(U+U^{\dag
}\right) \right] \right) .  \label{mat-mat}
\end{equation}%
Now, using Weyl's integration formula, and assuming that $M$ is a multiple
of the identity, that is, the masses of all quarks are taken identically $m$%
, one obtains \cite{LS} 
\begin{widetext}
\begin{equation}
Z_{\nu ,N_{f}}^{\mathrm{eff}}(m)=\frac{1}{(2\pi )^{N_{f}}N_{f}!}%
\int\limits_{-\pi }^{\pi }d\varphi _{1}\!\cdots \!\!\int\limits_{-\pi }^{\pi
}d\varphi _{N_{f}}\prod_{1\leq j<k\leq N_f}\left\vert \mathrm{e}^{i\varphi
_{k}}-\mathrm{e}^{i\varphi _{j}}\right\vert ^{2}\;\left(
\prod_{j=1}^{N_{f}}e^{V\Sigma m\cos (\varphi _{j})}\right) \left(
\prod_{j=1}^{N_{f}}e^{i\nu \varphi _{j}}\right).  \label{LS}
\end{equation}%
\end{widetext}
Notice that this matrix model (an integral of this type with a Vandermonde
term $\left\vert \mathrm{e}^{i\varphi _{k}}-\mathrm{e}^{i\varphi
_{j}}\right\vert ^{2}=4\sin ^{2}(\frac{\varphi _{k}-\varphi _{j}}{2})$ is a
unitary random matrix ensemble \cite{RM}), in the case $\nu =0$, is the
Gross-Witten matrix model, that appeared in the study of lattice 2d
Yang-Mills theory with the Wilson lattice action \cite{GW}\footnote{%
For $\nu \neq 0$ the matrix model describes the determinant of a Wilson loop
(P. Rossi, \href{http://www.sciencedirect.com/science/article/pii/0370269382908760}{Phys. Lett. B 117, 72 (1982)}) instead of the partition function of
the 2d Yang-Mills theory.}. In that theory there is no matter, so it
corresponds to two dimensional gluodynamics. Notice however a crucial
difference between the appearance of the matrix model in the two theories:
in the 2d Yang-Mills theory $N_{f}=0$\ and the group integration is then
over $U(N)$ which corresponds to $U(N_{c})$ in the 4d QCD case.\ Note that,
as pointed out in \cite{V-rev}, the matrix integration (\ref{mat-mat}) is
over $U(N_{f})$, the flavor space and hence, while the model is identical,
the description is very different. Taking this into account, we will also
focus on the very distinctive property of the Gross-Witten matrix model \cite%
{GW}: a third-order phase transition in the $N\rightarrow \infty $ limit which has
been the object of intense interest over three decades, since it plays a
rather paradigmatic role in the study of confinement/deconfinement and
Hagedorn phase transitions \cite{Aharony} and has also been a guide in the
study of phase transitions in 4d Yang-Mills theory \cite{Teper}. We shall
thus discuss both aspects of the correspondence with the spin chain: the
description of low-energy QCD in terms of the spin chain and the
implications of the Gross-Witten phase transition on the spin chain model.

\subsubsection{Complexity of the Leutwyler-Smilga integral}

Before proceeding to establishing and exploiting a spin chain representation of (\ref{qcd-theta}) and (\ref{LS}), we discuss some aspects of their numerical evaluation. In particular, by calling $\beta=V\Sigma m$, (\ref{LS}) is an integral representation of $\det \left( I_{i-j+\nu}\left( \beta \right) \right) _{i,j}^{n}$ where $I_{\nu }\left( \beta\right) $ denotes the modified Bessel function of second order, which is the 
$\nu $-th Fourier coefficient of the weight function of the matrix model, namely $e^{\beta \cos \theta }$.

The numerical evaluation of the Bessel function for a fixed small value of
the order $\nu $ is immediate since a numerical evaluation of its integral
representation with the trapezoidal rule is
exponentially convergent \cite{Tref}.

However, its evaluation for large values of the order $\nu $ and the
argument $\beta $ is a notoriously complex problem and the development of
numerical implementations of uniform asymptotic expansions of the Bessel
function in that regime is a subject of much current interest (see \cite{Jent} and
references therein). Indeed, even though the problem of its evaluation goes
back to Debye \cite{Deb}, who devised non-uniform asymptotic expansions, and
also that non-trivial uniform asymptotic expansions were found in the 1950s \cite{OlReprint,OlLoBoCl2010}, it turns out that the
coefficients of the asymptotic expansion not only exhibit 
resurgence \footnote{
Resurgence makes reference to the fact that the coefficients $a_{n}$ of a
Poincar\'{e} asymptotic series have the following special feature: the
coefficients in the asymptotic expansions of $a_{n}$, as $n\rightarrow
\infty $ are equal to themselves or related to the coefficients $a_{m}.$}
but also involve the evaluation of higher transcendental functions, in this
case Airy functions.

More specifically, the direct application of the uniform asymptotics
becomes problematic when the argument and the order of a Bessel function are
almost equal, due to huge numerical cancellations involved in
evaluating the individual coefficients in the uniform asymptotic expansions 
\cite{Jent} (which is a consequence of the confluence of two saddles in the steepest-descent study of the integral representation of the function).

The phenomena of the appearance and coalescence of saddles is of course
specially relevant when $\beta $ is complex, due to the different ensuing
crossings of Stokes lines in the steepest-descent study of the integral
representation of the Bessel function \cite{Berry}. 

At any rate, the numerical evaluation of (3) for a non-trivial topological
sector (i.e. very large $\nu $) is delicate at best because, in addition,
standard numerical implementations of the Bessel function can underflow for
large $\nu $ \cite{Amos}, in which case also the posterior evaluation of the
determinant with Gaussian elimination might be problematic.

This of course has the same implications for (\ref{qcd-theta}), where summing over all topological
sectors is involved. Notice also that it is possible to further characterize (\ref{qcd-theta}) analytically by plugging the integral representation (\ref{LS}) in (\ref{qcd-theta}), as was done in \cite{Lenaghan:2001ur}. However, the resulting expression, as expected, loses its random matrix/determinantal form and it involves the
evaluation of Bessel functions \emph{and} a posterior multivariable
integration with the same weight as in (\ref{LS}), but with the Bessel functions in
the integrand \cite{Lenaghan:2001ur}. Only in the two simplest cases, corresponding to one and two
flavours (one and two spins flipped, in our forthcoming picture) has the
partition function an explicit analytic expression \cite{Lenaghan:2001ur}.

The connection made in this paper (equation (\ref{keyequation}) below) opens the possibility of using numerical methods developed in the study of spin chains, such as White's Density Matrix Renormalization Group (DMRG) algorithm \cite{White}, or more concretely some of its finite-temperature versions like \cite{T-DMRG}, as an alternative method to compute the Leutwyler-Smilga integral for real $\beta$. For imaginary $\beta$, where classical simulation methods usually break down for large $\beta$, one may use quantum simulations with optical lattices. Indeed, as we comment below when discussing experimental implementations, the experiment \cite{Fukuhara} does exactly the job.

\subsection{1D XX model and thermal correlation functions}

Let us now describe the result in \cite{Bog,Bog2,B,BC} which relates some
thermal correlation function of the XX model to a matrix model, which turns
out to be the same as before.

Let us begin our discussion by presenting the spin chain model. The $S=1/2$
Heisenberg XX spin chain is one of the simplest integrable magnetic chains.
It has a well-known mapping, using the Jordan-Wigner transformation, to a
free fermion system \cite{LSM}. This infinite chain (which we consider with periodic boundary conditions) is characterized by the Hamiltonian%
\begin{equation}
\hat{H}=-\frac{1}{2}\sum_{i}\sigma _{i}^{-}\otimes \sigma _{i+1}^{+}+\sigma
_{i}^{-}\otimes \sigma _{i-1}^{+}+\frac{h}{2}\sum_{i}(\sigma _{i}^{z}-%
\mathbbm{1}),  \label{XX-o}
\end{equation}%
where the summation is over all lattice sites and $h>0$. As usual, $\sigma _{i}^{\pm
}=\left( \sigma _{i}^{x}\pm i\sigma _{i}^{y}\right) /2,$ where $\sigma
_{i}^{x}$ and $\sigma _{i}^{y}$ together with $\sigma _{i}^{z}$ denote the
Pauli spin operators and $h$ represents the strength of an external magnetic
field. The commutation relations are%
\begin{equation*}
\lbrack \sigma _{i}^{+},\sigma _{k}^{-}]=\sigma _{i}^{z}\delta _{ik},\qquad
\lbrack \sigma _{i}^{z},\sigma _{k}^{\pm }]=\pm 2\sigma _{i}^{\pm }\delta
_{ik}.
\end{equation*}%
These operators are nilpotent $(\sigma _{i}^{\pm })^{2}=0$, a property that
will lead to a determinantal form for the correlation functions that we
shall focus on. The other operator satisfies $(\sigma _{i}^{z})^{2}=1$.

Let us begin by defining and describing the correlation functions of the
model. Thermal correlation functions of spin chains have been studied for
some time \cite{T1} and are known to admit determinantal expressions which
are simpler in the case of the XX model (\ref{XX-o}) and have been studied
explicitly more recently \cite{Bog,Bog2,BC,B}. Following \cite{B}, the
correlation function will be defined on a ferromagnetic state, which is
characterized by having all the spins up ${}\,\lvert \,\Uparrow \,\rangle=\otimes _{i}{}\,\lvert \,\uparrow \,\rangle _{i}$, which satisfies $\sigma_{k}^{+}{}\,\lvert \,\Uparrow \,\rangle =0$ for all $k,$ and the state is
also normalized $\langle \,\Uparrow \,\mid \,\Uparrow \,\rangle =1$. This
state is annihilated by the Hamiltonian $\hat{H}{}\,\lvert \,\Uparrow\,\rangle =0$ and the thermal correlation functions are defined by 
\begin{equation*}
F_{j_{1},\dots ,j_{K};l_{1},\dots ,l_{K}}(\beta )=\langle \,\Uparrow
\,\rvert \,{}\sigma _{j_{1}}^{+}\cdots \sigma _{j_{K}}^{+}\mathrm{e}^{-\beta 
\hat{H}}\sigma _{l_{1}}^{-}\cdots \sigma _{l_{K}}^{-}{}\,\lvert \,\Uparrow
\,\rangle .
\end{equation*}%
\newline 
Let us consider first the particular case where we only have one spin down $%
K=1.$ 
\begin{equation}
F_{jl}(\beta )=\langle \,\Uparrow \,\rvert \,{}\sigma _{j}^{+}\mathrm{e}%
^{-\beta \hat{H}}\sigma _{l}^{-}{}\,\lvert \,\Uparrow \,\rangle \; . \label{F}
\end{equation}%
By taking into
account the commutation relation%
\begin{equation}
\lbrack \sigma _{j}^{+},\hat{H}]=-\frac{1}{2}\sum_{k}\Lambda _{jk}\sigma
_{j}^{z}\sigma _{k}^{+}-h\sum_{k}\sigma _{k}^{+}\delta _{j,k}  \label{comm}
\end{equation}%
\begin{equation*}
=-\frac{1}{2}\sigma _{j}^{z}(\sigma _{j-1}^{+}+\sigma _{j+1}^{+})-h\sigma
_{j}^{+},
\end{equation*}%
where $\Lambda _{jk}=\delta _{j,k+1}+\delta _{j,k-1}$, together with the
property $\langle \,\Uparrow \,\rvert \,{}\sigma _{j}^{z}=\langle \,\Uparrow
\,\rvert \,{}$, it follows immediately \cite{B}%
\begin{equation*}
\frac{d}{d\beta }F_{jl}(\beta )=-\langle \,\Uparrow \,\rvert \,{}\sigma
_{j}^{+}\hat{H}\mathrm{e}^{-\beta \hat{H}}\sigma _{l}^{-}{}\,\lvert
\,\Uparrow \,\rangle 
\end{equation*}%
\begin{equation*}
=\frac{1}{2}\langle \,\Uparrow \,\rvert \,{}(\sigma _{j-1}^{+}+\sigma
_{j+1}^{+}+2h\sigma _{j}^{+})\mathrm{e}^{-\beta \hat{H}}\sigma
_{l}^{-}{}\,\lvert \,\Uparrow \,\rangle .
\end{equation*}%
Hence%
\begin{equation}
\frac{d}{d\beta }F_{jl}(\beta )=\frac{1}{2}\left( F_{j+1,l}(\beta
)+F_{j-1,l}(\beta )\right) +hF_{j,l}(\beta ).  \label{rw}
\end{equation}%
This equation is that of a symmetric random walk on a line. Let us also remark that by commuting $\hat{H%
}$ with $\sigma _{l}^{-}$, there is an analogous difference equation but for
subscript $l$ with fixed subscript $j$ \cite{B}. Both equations are subject
to the initial condition $F_{jl}(0)=\delta _{jl}$, and to boundary
conditions that depend on the type of lattice considered. The results in \cite{B}
show that the case of general $K>1$ generalizes in a straightforward way and
the multi-dimensional analogue of (\ref{rw}) is obtained. The initial
condition is the same $F_{j_{1},\dots ,j_{K};l_{1},\dots ,l_{K}}(0)=\delta
_{j_{1}l_{1}}\cdots \delta _{j_{K}l_{K}}$ and the correlation function also
satisfies the conditions $F_{j_{1},\dots ,j_{K};l_{1},\dots ,l_{K}}(\beta )=0
$ if $l_{r}=l_{s}$ or $j_{r}=j_{s}$ ($r,s=1,\dots ,K$), due to the
nilpotency of the spin operators, $(\sigma _{i}^{\pm })^{2}=0$. This
"non-intersecting" property suggests a determinantal structure and indeed,
the solution of the equation for general $K$ can be expressed as \cite{B}%
\begin{equation}
F_{j_{1},\dots ,j_{K};l_{1},\dots ,l_{K}}(\beta )=\det_{1\leq r,s\leq
K}\left\{ F_{j_{r}l_{s}}(\beta )\right\} ,  \label{minordet}
\end{equation}%
where $F_{jl}(\beta)$ are the one-particle correlation functions satisfying (\ref%
{rw}). A matrix model expression for this determinant is given by \cite%
{B,AvM,Minor}%
\begin{widetext}
\begin{equation}
F_{j_{1},\dots ,j_{K};l_{1},\dots ,l_{K}}(f,\beta )=\frac{1}{(2\pi )^{K}K!}%
\int\limits_{-\pi }^{\pi }d\varphi _{1}\!\cdots \!\!\int\limits_{-\pi }^{\pi
}d\varphi _{K}\prod_{1\leq j<k\leq K}\left\vert \mathrm{e}^{i\varphi _{k}}-%
\mathrm{e}^{i\varphi _{j}}\right\vert ^{2}\;\left( \prod_{j=1}^{K}f(\varphi
_{j})\right) \overline{\hat{s}_{\alpha }\left( e^{i\varphi _{1}},\dots
,e^{i\varphi _{K}}\right) }\hat{s}_{\gamma }\left( e^{i\varphi _{1}},\dots
,e^{i\varphi _{K}}\right) ,  \label{mat}
\end{equation}%
\end{widetext}where $\hat{s}_{\lambda }\left( e^{i\varphi _{1}},\dots
,e^{i\varphi _{n}}\right) $ is a Schur polynomial, a symmetric polynomial 
\cite{Stanley}. The relationship between the partitions $\alpha $ and $\gamma $ 
in the r.h.s. of (\ref{mat}) and the $j$ and $l$ that appear in the thermal 
correlation function is \cite{B} 
\begin{eqnarray*}
\alpha _{r} &=&j_{r}-K+r \\
\gamma _{r} &=&l_{r}-K+r
\end{eqnarray*}%
and the weight function $f(\varphi )$ in the matrix model (\ref{mat}) is the generating function of the one-spin flip
process (\ref{rw}). Therefore, noticing that (\ref{mat}) is of the same form as (\ref{LS}) but more general due to 
the presence of two Schur polynomials, we can identify the number of flipped spins $K$ with the number of flavors $N_f$. 
In addition, $f(\varphi )$ being the generating function of the one-spin flip
process (\ref{rw}), is generically given by 
\begin{equation*}
f\left( \beta ,e^{i\varphi }\right) =\sum_{j=-\infty }^{\infty }F_{jl}(\beta
)e^{i\varphi j},
\end{equation*}%
and, in this particular case, by%
\begin{equation}
f\left( \beta ,e^{i\varphi }\right) =f\left( 0,\lambda \right) \exp \left( 
\frac{\beta }{2}\left( 2h+(e^{i\varphi }+e^{-i\varphi })\right) \right) .
\label{genfunc1}
\end{equation}%
Therefore, the weight function in (\ref{mat}) is%
\begin{equation}
f\left( \varphi \right) =\mathrm{e}^{\beta (h+\cos \varphi )}.
\label{weight}
\end{equation}
Notice that the analysis and the final result for the case of one flipped
spin $K=1$ is identical to Glauber's seminal study of the kinetic Ising
model \cite{Glauber}. In particular, the thermal
average (\ref{F}) behaves like the expectation of a single spin in an infinite
ring in \cite{Glauber}, with the time variable in \cite{Glauber} identified
with our $\beta $. 

By considering now the specific pattern of flipped spins $j_{r}=\nu +K-r$ and $l_{r}=K-r$, we get in (\ref{mat})
the partitions $\alpha =(0,\ldots ,0)$ and $\gamma =(\nu ,\ldots ,\nu )$.
Using that in this case \cite{Stanley} 
\begin{equation*}
\hat{s}_{\alpha }\left( e^{i\varphi _{1}},\dots ,e^{i\varphi _{K}}\right) =1
\end{equation*}%
\begin{equation*}
\hat{s}_{\gamma }\left( e^{i\varphi _{1}},\dots ,e^{i\varphi _{K}}\right)
=\prod_{j=1}^{K}e^{i\nu \varphi _{j}},
\end{equation*}%
we recover exactly equation (\ref{LS}) from equation (\ref{mat}). That is,
we obtain the key equation%
\begin{equation}
\langle ...,\uparrow ,\underset{N_{f}}{\underbrace{\downarrow
,...,\downarrow }},\underset{\nu }{\underbrace{\uparrow ,...,\uparrow }}%
\rvert \mathrm{e}^{-\beta \hat{H}_{\mathrm{XX}}}\lvert \underset{N_{f}}{%
\underbrace{\downarrow ,\downarrow ,...,\downarrow }},\uparrow ,...\rangle
=Z_{\nu ,N_{f}}^{\mathrm{eff}}(m)  \label{keyequation}
\end{equation}
where $\beta=V\Sigma m$.

\subsection{A dictionary QCD -- spin chains}

\subsubsection{The topological sector}

The first obvious connection arising from equation (\ref{keyequation}) is
that the number of flavors $N_f$
corresponds to the number of spins down or, equivalently in the free fermion
picture, the particle-number sector to which we restrict our attention in
the 1D spin chain.

On top of that, the shift $\nu $ in the positions of the spin down particles
at both sides of the thermal average in equation (\ref{keyequation}) induces
a phase change in equation (\ref{LS}), responsible for the non-trivial
topological sector of the QCD partition function. How to understand this as some type of topological order present in the XX spin chain? The question is tricky since there is in principle no clear way to define topological order in 1D. A possible answer comes from the holographic principle, where one sees a (not necessarily normalized) 1D thermal state $e^{-\beta H}$ as the {\it  boundary} of a 2D system. If the 2D system is topologically ordered, this should leave some signature in the 1D state. Starting with the seminal work of Li and Haldane \cite{Haldane}, there has been several recent discussions about which this signature is \cite{Qi, Schuch1, Schuch2, HongHao}. Two key facts can be extracted from there: (i) each topological sector in the bulk corresponds to projecting the thermal state of the boundary Hamiltonian in a different sector; and (ii) the bulk topology  translates to some {\it dynamical property} on the boundary, and is hence related to the momentum. This agrees with the appearance of the translation operation $T$ in equation (\ref{keyequation}), which can be simply restated as:
$$
Z_{\nu ,N_{f}}^{\mathrm{eff}}(m)=\langle ...,\uparrow ,\underset{N_{f}}{\underbrace{\downarrow
,...,\downarrow }}%
\rvert \mathrm{e}^{-\beta \hat{H}_{\mathrm{XX}}}T^{-\nu}\lvert \underset{N_{f}}{%
\underbrace{\downarrow ,...,\downarrow }},\uparrow ,...\rangle
$$
$$=\langle ...,\uparrow ,\underset{N_{f}}{\underbrace{\downarrow
,...,\downarrow }}%
\rvert \mathrm{e}^{-\frac{\beta}{2} \hat{H}_{\mathrm{XX}}}T^{-\nu} \mathrm{e}^{-\frac{\beta}{2} \hat{H}_{\mathrm{XX}}} \lvert \underset{N_{f}}{%
\underbrace{\downarrow ,...,\downarrow }},\uparrow ,...\rangle.
$$
The last equation resembles very much the momentum polarization tool introduced very recently in \cite{HongHao} as a way to
detect non-trivial topological behavior \footnote{Following \cite{HongHao}, one can see how (\ref{keyequation}) relates directly with the generator $L_0$ of the Virasoro algebra associated with the CFT of the XX-model. This emphasizes, in a way similar to \cite{HongHao}, the topological content of $\nu$}. 

To get (i) and (ii) and then show in a clearer way the topological content of $\nu$ it is better to go back to its Fourier dual parameter $\theta$.  In order to avoid unnecessary mathematical complications we will assume now a finite chain of $2L+1$ spins and define $\hat{T}=\frac{1}{2L+1}\sum_{\nu=-L}^Le^{i\nu\theta} T^{-\nu}$. By taking a basis of states $|k\rangle$ with definite momentum $T|k\rangle=e^{i\frac{2\pi k}{2L+1}}|k\rangle$  and changing variables $\theta=\frac{2\pi \theta'}{2L+1}$ one can see that 
$$\hat{T}|k\rangle =\frac{1}{2L+1}\sum_{\nu=-L}^Le^{\frac{2\pi \nu i}{2L+1} (\theta'-k)}|k\rangle =\delta_{k,\theta'} |k\rangle$$
and hence $\hat{T}$ is just the projector $P_\theta$ onto the states with momentum $\theta$. Since it commutes trivially with the Hamiltonian, we get finally
$$\frac{1}{2L+1}\sum_{\nu=-L}^Le^{i\nu\theta} \langle ...,\uparrow ,\underset{N_{f}}{\underbrace{\downarrow
,...,\downarrow }}%
\rvert \mathrm{e}^{-\beta \hat{H}_{\mathrm{XX}}}T^{-\nu}\lvert \underset{N_{f}}{%
\underbrace{\downarrow ,...,\downarrow }},\uparrow ,...\rangle
 \;$$
 $$
 = \langle ...,\uparrow ,\underset{N_{f}}{\underbrace{\downarrow
,...,\downarrow }}%
\rvert P_\theta \mathrm{e}^{-\beta \hat{H}_{\mathrm{XX}}}P_\theta\lvert \underset{N_{f}}{%
\underbrace{\downarrow ,...,\downarrow }},\uparrow ,...\rangle.$$
By considering the limit $L\rightarrow \infty$ and  (\ref{qcd-theta}) this has the extra benefit of giving an interpretation of the global partition function $Z_{QCD}^{\mathrm{eff}}(\theta)$ as a thermal average on the XX-model when the Hamiltonian $\hat{H}$ is projected onto the sector of momentum $\theta$. That is,
$Z_{QCD}^{\mathrm{eff}}(\theta)=$
\begin{equation}\label{eq:theta}
\lim_{L\rightarrow\infty} (2L+1) \langle \overset{2L+1} {\overbrace{...,\uparrow ,\underset{N_{f}}{\underbrace{\downarrow
,...,\downarrow }}}}%
\rvert P_\theta \mathrm{e}^{-\beta \hat{H}_{\mathrm{XX}}}P_\theta\lvert \underset{N_{f}}{%
\underbrace{\downarrow ,...,\downarrow }},\uparrow ,...\rangle\;  . 
\end{equation}
A mathematically fully rigorous argument of that will be provided in the Appendix.

\subsubsection{Different matter content}

In the random matrix description of the thermal correlators one can obtain
symmetries other than the unitary symmetry of (\ref{mat}). As happens with
the analogous setting of the Calogero model \cite{per} and of
non-intersecting random walks \cite{Grab,Greg}, the inclusion of boundaries
in the problem leads to other symmetries, such as orthogonal and symplectic
symmetries. One of these cases is actually treated explicitly in \cite{B},
where an absorbing boundary condition at the origin is shown to lead to the
same matrix model, but with a correlation term between eigenvalues%
\begin{equation}
\prod_{i=1}^{K}\sin ^{2}\theta _{i}\prod_{1\leq j<k\leq K}\sin ^{2}\left( 
\frac{\theta _{j}-\theta _{k}}{2}\right) \sin ^{2}\left( \frac{\theta
_{j}+\theta _{k}}{2}\right)   \label{symplectic}
\end{equation}%
instead of the usual Vandermonde in (\ref{mat}). These other situations have
a counterpart in the low-energy QCD. In the chiral limit (with the masses of
the fermions $m_{f}\rightarrow 0$) the relevant random matrix ensembles are
the chiral GUE, chiral GOE and chiral GSE ensembles \footnote{%
Gaussian orthogonal, unitary and symplectic ensembles, respectively \cite{RM}%
.} \cite{V}, which are the ensembles that appear when the gauge theory has $%
SU(N_{c})$ symmetry, with $N_{c}\geq 3$, for $SU(2)$ gauge group, and again
for $SU(N_{c})$ and $N_{c}\geq 3$ but in the adjoint representation (and
fermions in the adjoint representation are Majorana fermions \cite{LS}),
respectively \cite{V}. These are precisely the resulting ensembles that
describe the spin chain in the limit $\beta \rightarrow \infty $, because
the weak-coupling limit of the Gross-Witten model is a Gaussian unitary
ensemble \cite{RT}.
The limitation in this case is due to the fact that 
$\beta$ is the parameter in the weight function of the resulting Gaussian 
ensemble and therefore, taking into account the identification of parameters 
in \cite{V}, this implies that there is a corresponding vanishing limit of the 
quark condensate. Thus, the correspondence in this setting is more subtle, due to the
role played by Gaussian ensembles, which only emerge in our setting in the limit 
$\beta \rightarrow \infty $. Hence, these other cases have to be considered in more detail 
but the point is that other relevant symmetries can be in principle described by 
considering boundaries in the spin chain model.

\subsubsection{Effects of a lattice}

In addition to other symmetries, obtained with the inclusion of boundaries
in the spin chain, one can also consider additional interactions between
neighboring spins in the chain. These new interactions modify accordingly
the weight function in the matrix model (\ref{mat}). This allows to extend
the correspondence between the spin chain and low-energy QCD to the case
where the gauge theory is studied on the lattice \cite{DSV,ADSV}. The
lattice breaks the chiral symmetry explicitly and hence the effects of the
lattice spacing lead to new terms in chiral perturbation theory. This
extended low energy theory is known as Wilson chiral perturbation theory and
leads to an extension of the matrix model (\ref{mat-mat}), characterized by
the addition of potential terms \cite{DSV,ADSV} 
\begin{equation*}
V(U)=-a^{2}VW_{6}\mathrm{Tr}\left[ \left( U+U^{\dag }\right) \right]
^{2}-a^{2}VW_{7}\mathrm{Tr}\left[ \left( U-U^{\dag }\right) \right] ^{2}
\end{equation*}%
\begin{equation}
-a^{2}VW_{8}\mathrm{Tr}\left( U^{2}+U^{\dag 2}\right) ,  \label{V}
\end{equation}%
where $a$ denotes the lattice spacing and $W_{6},W_{7}$ and $W_{8}$ are the
new low energy constants. The first two terms in (\ref{V}) are multi-trace
potentials which are more difficult to treat in general and, for the moment,
have no known spin chain representation. However, these terms are expected to
be suppressed in the large $N_{c}$ limit and are often not considered \cite{DSV,ADSV}.
Interestingly enough, the remaining potential term in (\ref{V}) can be
described in the same manner as above, just by generalizing the spin chain
to include next-to nearest neighbors interactions. The resulting Hamiltonian
is then%
\begin{equation*}
\hat{H}=-\frac{1}{2}\sum_{i}J_{1}\left( \sigma _{i}^{-}\otimes \sigma
_{i+1}^{+}+\sigma _{i}^{-}\otimes \sigma _{i-1}^{+}\right) +
\end{equation*}%
\begin{equation*}
+J_{2}\left( \sigma _{i}^{-}\otimes \sigma _{i+2}^{+}+\sigma _{i}^{-}\otimes
\sigma _{i-2}^{+}\right) .
\end{equation*}%
Notice that we previously have identified the $\beta $ parameter of the spin
chain with a single combination of parameters of the effective field theory: 
$\beta =mV\Sigma $. Now we have to identify $\beta J_{1}=mV\Sigma $ and $%
\beta J_{2}=2a^{2}VW_{8}$. Thus, the relative strength of the interactions at
first and second neighbors depends on the quotient between the masses of the
quarks and the lattice spacing, together with the respective low-energy
constants%
\begin{equation*}
\frac{J_{1}}{J_{2}}=\frac{m\Sigma }{2a^{2}W_{8}}.
\end{equation*}

\subsection{Finite chain errors. Experimental accessibility}\label{sec:finite}

It is also shown in \cite{BC}, with a similar argument, that in the case of
a finite chain of $L$ sites, the thermal average at the right hand side of
equation (\ref{keyequation}) is nothing but the Riemann sum associated to
the integral (\ref{LS}) when we evaluate on the vertices of a lattice
division of the hypercube $[-\pi ,\pi ]^{N_{f}}$ of length $\frac{2\pi }{L}$%
. A recent result by Baik and Liu \cite{BaikLiu} shows that the error obtained  by this
particular Riemann sum approximation decreases exponentially with $L$. More concretely, the relative error is $O(e^{-c(L-N_f)})$ as $L-N_f\rightarrow \infty$, even if $N_f$ also goes to $\infty$. 

This opens the door to a possible experimental measure of the quantity $%
Z_{\nu ,N_{f}}^{\mathrm{eff}}(\beta)$ as long as the experimental setup allows the following four steps:  (1) implement the XX-Hamiltonian $\sum_i \sigma_i^+\otimes \sigma_{i+1}^- +h.c.$ (preferably with a magnetic field on the $z$-direction),  (2) enforce
the sector of $N_f$ particles, (3) stabilize the thermal state within the
sector and (4) measure the positions where the particles are. The crucial point is to realize that then the size
of the chain, and the number of times that the experiment has to be done in
order to approximate $Z_{\nu ,N_{f}}^{\mathrm{eff}}(\beta )$ within a relative error 
$\epsilon $ scales only polynomially with $\log(\frac{1}{\epsilon })$ (we treat $%
N_{f}$ and $\beta $ as constants). In order to see that, we consider a
magnetic field $h\leq -2$ such that $|\Uparrow \rangle $ is the ground
state. Note that by equation (\ref{weight}), the magnetic field only gives a
factor $e^{\beta hN_{f}}$ in the thermal average so we can choose its value
to our convenience. By the bound on the relative error, the length of the chain needs to
scale only linearly with $\log(\frac{1}{\epsilon })$. The first step is to
restrict to the sector given by $N_{f}$ spins down (particles in the free
fermion picture). In this way, one gets a Hilbert space $\mathcal{H}_{N_f}$,
whose dimension scales polynomially with $L$, and hence with $\log(\frac{1}{%
\epsilon })$. The next step is to stabilize the system at the desired
temperature, obtaining the thermal state $\rho _{\beta }$ which is nothing
but $\frac{e^{-\beta \tilde{H}}}{\mathrm{tr}\,e^{-\beta \tilde{H}}}$ being $%
\tilde{H}$ the restriction of $\hat{H}$ to $\mathcal{H}_{N_f}$. Since we have
chosen the magnetic field for the state $|\Uparrow \rangle $ (the vacuum in
the free fermion picture) to be the ground state, it is not difficult to see
that $\mathrm{tr}\,e^{-\beta \tilde{H}}\leq \dim \mathcal{H}_{N_f}$,
which makes 
\begin{equation}
\left\vert \langle \Uparrow |\sigma _{1}^{+}\cdots\sigma
_{N_f}^{+}\rho _{\beta }\sigma _{N_f}^{-}\cdots\sigma _{1}^{-}|\Uparrow
\rangle \right\vert \geq \frac{1}{\mathrm{polylog}(\frac{1}{\epsilon })}
\label{eq-free-fermions-3}
\end{equation}

But the lefthand side of (\ref{eq-free-fermions-3}) is the probability of,
given the state $\rho_\beta$ and measuring where the three particles are,
obtaining that they are in positions one to three. Since this is larger than
$\frac{1}{\mathrm{polylog}(\frac{1}{\epsilon })}$, the number of times one needs to make
the experiment in order to get this value accurately scales also
polynomially with $\log(\frac{1}{\epsilon})$.

It seems that ultra cold gases in optical lattices are the best system nowadays to get the required steps (1)-(4). Indeed, very recently \cite{Fukuhara}, the quantity $Z_{\nu ,N_{f}}^{\mathrm{eff}}(\beta)$ have been measured for chains of around 20 sites with imaginary $\beta$. The case of real $\beta$ does not seem completely out of reach.  Let us briefly discuss why. There are two ways of getting the XX-Hamiltonian in an optical lattice. 
One is to implement a 1D lattice Hard Core Boson Hamiltonian which, by considering the particle-hole degree of freedom, is exactly the XX-Hamiltonian. 
This (with an extra periodic confining potential)  was already shown experimentally in \cite{Paredes}, getting an array of 1D systems with a probability  greater than $\frac{1}{11M^\frac{1}{3}}$ of having at least one with $M$ particles (for $M$ small).
A different route to get such Hamiltonian, proposed before in \cite{Duan} and  experimentally obtained in \cite{Fukuhara}, is to consider atoms with a spin degree of freedom in the insulating phase. By tuning appropriate the parameters, in second-order perturbation theory one obtains the XX-Hamiltonian as the effective Hamiltonian of the system, though in a much smaller energy scale. By using the single spin addressing recently developed in \cite{Weitenberg}  as done in \cite{Fukuhara} one can enforce the sector of a fixed number of particles.  By the recent technique of high-resolution fluorescence imaging \cite{Greiner,Bloch}, one may also measure the position of the particles. The most subtle issue is stabilizing the thermal state. Indeed, the understanding of the thermodynamical properties of ultracold gases in optical lattices is a hot topic nowadays \cite{McKay}, which may lead to a solution of this problem in the near future. There are at least two possible routes for that \cite{McKay}. One may start with a Bose-Einstein Condensate (BEC) in thermal equilibrium which is then adiabatically loaded into the lattice potential. Even though the XX-Hamiltonian is integrable and, as shown for instance in \cite{cradle, Rigol} thermalization without an external bath is not guaranteed, the measures made in \cite{Paredes} are in excellent agreement with having a thermal state \cite{Paredes, Rigol2}. A second approach is to immerse the system in a reservoir with particles of a different species \cite{McKay}, so that we keep the number of particles constant in the lattice. Though there is no full study of the expected thermalization time, the recent estimate of \cite{Temme} for the gap of the Davies Liouvillian -- the one modeling the convergence to the thermal state of a system weakly coupled to a thermal bath-- in the case of a fermion hoping on a line, allows one to be  optimistic in this direction.

\subsection{A third order phase transition on the XX chain and the Tracy-Widom law}

The final implication of (\ref{keyequation}) is the existence of a third
order phase transition hidden in the XX-model \footnote{Third order phase transitions are rare in condensed matter systems. Another third order phase transition was found recently in \cite{Ortiz}} -- the so called Gross-Witten
transition. Let us recall here that, as was shown in the seminal paper \cite%
{GW}, if we consider the t'Hooft parameter $\lambda =K/\beta$, and we make $%
K\rightarrow \infty $ while keeping $\lambda $ constant, we obtain a
double-scaling limit in $Z_{\nu =0,K}(\frac{\beta}{V\Sigma} )$ --now we call it $Z_{\mathrm{%
GW}}(\beta ,U(K))$ since it is the partition function of the Gross-Witten
model with gauge group $U(K)$-- with a third-order phase transition between
the two regimes. Formally, the limit for the free energy 
\begin{equation*}
F_{K}\left( \lambda \right) =\frac{1}{K^{2}}\ln Z_{\mathrm{GW}}(\beta ,U(K))
\end{equation*}
gives us \cite{GW} 
\begin{equation*}
\lim_{K\rightarrow \infty}F_{K} (\lambda) =\left\{ 
\begin{array}{ll}
\frac{1}{4\lambda^{2}} & ,\lambda \geq 1 \\ 
\frac{1}{\lambda }+\frac{1}{2}\ln \lambda -\frac{3}{4} & , \lambda <1%
\end{array}
\right.
\end{equation*}
By (\ref{keyequation}), which now reads 
\begin{equation}
\langle ...,\uparrow ,\underset{K}{\underbrace{\downarrow ,...,\downarrow
,\downarrow }}\rvert \mathrm{e}^{-\beta \hat{H}_{\mathrm{XX}}}\lvert 
\underset{K}{\underbrace{\downarrow ,\downarrow ,...,\downarrow }},\uparrow
,...\rangle =Z_{\mathrm{GW}}(\beta ,U(K))  \label{flipped}
\end{equation}
this is a phase transition in the XX model. Notice that now the correspondence is between the number of flipped spins and the rank of the gauge group. 
It is noteworthy that the above mentioned exponentially small error for chains of finite size $L$ 
also holds in the double-scaling limit \cite{BaikLiu,Baik2}. In particular, for the two phases, it holds that, if $R(L,K,\beta)$ denotes the Riemann sum, which is the partition function of the finite size spin chain, divided by the multiple integral (the partition function of the infinite chain), then \cite{Baik2} $R(L,K,\beta) = 1 +O(e^{-cK})$    if $L> (1+\epsilon) \mu(K,\beta)$ (with $\epsilon >0$), where
\begin{equation}\label{eq:msca}
    \mu(K,\beta):= \begin{cases}
    2\sqrt{K\beta}, \qquad & \lambda<1,\\
    K+\beta, \qquad &\lambda\ge 1.
    \end{cases}
\end{equation}
Note the different scaling of the finite length $L$ in terms of the number of flipped spins and the inverse temperature, depending on the phase. This property
is a direct consequence of the discreteness of the associated random matrix ensemble.

On the other hand, one may argue that this phase transition only happens in a very unnatural limit of the spin chain
parameters. However, there are signatures of this phase transition (a
crossover) for very small values of $K$, as can be seen from the plots for $K=1,2,3$ in \cite{crossover}. Actually, stronger results are available, since the following estimate holds in the strong-coupling phase ($\lambda\ge 1$)  \cite{BDJ}

\begin{equation}
\left\vert F_{K}\left( \lambda \right)- F\left( \lambda \right)\right\vert \leq Ce^{-cK},
\end{equation}
for some constants $C$ and $c$ and $F\left( \lambda \right)=\lim_{K\rightarrow \infty}F_{K} (\lambda)$. This predicts a exponentially small departure, in the strong-coupling phase, for the case of a finite number of flipped spins $K$. 

Notice that this result holds for the free energy, which is the logarithm of the thermal correlator, and not just the partition function. The other phase is a bit more
delicate to analyze and it is known that the finite rank case has $1/K$ and higher-order corrections \cite{Goldschmidt:1979hq}. The ressummation of 
all the infinitely many terms is of course a complex and delicate issue, but more recent results show that the departure with the infinite rank case decays 
quickly also in the weak-coupling phase \cite{Marino:2008ya}.

Thus, taking into account the results on finite chain errors and the comments above on experimental accessibility one may be able to
observe the Gross-Witten phase transition experimentally in a spin chain. 

Besides the work of Gross and Witten, the random matrix ensemble (\ref{LS}) with $\nu =0$ is central in the
ground breaking description of the asymptotics of the length of the longest
increasing subsequence in random permutations \cite{BDJ}. In \cite{BDJ}, it was proved
that writing%
\begin{equation}
K=\beta +x\left( \beta /2\right) ^{1/3}  \label{scaling}
\end{equation}%
and for $\beta \rightarrow \infty $, then%
\begin{equation}
\mathrm{e}^{-\beta ^{2}/4}Z_{GW}\left( \beta ,U(K)\right) \rightarrow
F_{2}(x),  \label{TW}
\end{equation}%
where $F_{2}(x)$ is the celebrated Tracy-Widom distribution \cite{TW} which
can be given in terms of the Fredholm determinant of an Airy kernel or
through an integral representation involving a solution of the Painlev\'{e}
II equation, from which asymptotic expansions for $F_{2}(x)$ follow \cite%
{TW,RM}. 
Notice that the factor $\mathrm{e}^{-\beta ^{2}/4}$ in (\ref{TW}) is the
normalization constant of the matrix model in the limit $K\rightarrow \infty 
$. In this way the l.h.s. of (\ref{TW}) is actually 
\[
F_{\left\{ K-r\right\} _{r=0}^{K-1},\left\{ K-r\right\} _{r=0}^{K-1}}\left(
\beta \right) \text{ with }\lim_{K\rightarrow \infty }F\left( \beta,h^{\ast
} \right)
=1\text{.}
\]%

If the spin chain does not have the magnetic field term in (\ref{XX-o}) then one
has to add the prefactor in (\ref{TW}) by hand. With the magnetic field $h$
there is an overall prefactor $\exp \left( h\beta K\right) $ outside
the matrix model integral representation and, hence, to observe the Tracy-Widom shape
which emerges in the scaling region above, one can fine-tune the magnetic
field to $h^{\ast}=\frac{\beta }{4K}\sim 1/4$.

Notice that the result above zooms in the $\lambda =1$ region where the 3rd
order phase transition occurs, and it describes typical and small
fluctuations of order $\mathcal{O}\left( K^{-2/3}\right) $ in the transition
point \cite{FMS}\footnote{%
Notice that we can write the condition (\ref{scaling}) as $x=2^{1/3}\frac{%
\lambda -1}{K^{2/3}}$, where $\lambda $ is the 't Hooft parameter.}. The
relevance of the Tracy-Widom distribution resides on its large
universality and the universal fluctuations that it characterizes have been measured in recent experiments studying the height distribution of interfaces, in particular in
the slow combustion of paper and in turbulent liquid crystals \cite%
{Exp1,Exp2}. Note that it makes its appearance in 2d classical statistical
mechanics systems \cite{rev-TW} whereas we are proposing that a corresponding result also
holds in a quantum 1d system, the Heisenberg XX model.

\subsection{Conclusions}

We have uncovered a connection between QCD and the XX model using random
matrix models which allows to establish a dictionary between both worlds as
sketched in Table \ref{tab:dictionary}. This opens an avenue to connect
different QFT with 1D spin chain Hamiltonians. Specially interesting is the
case of Chern-Simons theory due to its connections with topology, knot
theory and the fractional quantum Hall effect. We will make a full study
in a forthcoming paper, showing that Table \ref{tab:dictionary} also
applies, albeit with a different 1D spin Hamiltonian.

We thank Gernot Akemann, Mari Carmen Ba\~nuls,  Ignacio Cirac, Daniel Fern\'andez-Fraile, Juan Jos\'e Garc\'{\i}a-Ripoll, Jose Ignacio Latorre, Bel\'en Paredes, Gr\'egory Schehr, Germ\'an Sierra, and Pere Talavera for many useful comments and suggestions. We are indebted to Jinho Baik for very precise explanations of the results in \cite{BaikLiu}. This work has been partially funded by Spanish grants MTM2011-26912 and
QUITEMAD, Spanish program Juan de la Cierva and European CHIST-ERA project CQC.

\appendix

\subsection{Appendix:\ Proof of (\ref{eq:theta})}

We will use some simple  Banach space tools and notations for that. Let us recall that $\ell_1$ is the Banach space of sequences $x=(x_n)_{n\in \Z}$ such that $\|x\|_1=\sum_{n=-\infty}^\infty |x_n|<\infty$. It is clear that the direct sum (or cartesian product) of a finite number of copies of $\ell_1$, $\oplus_{k=1}^K\ell_1$, can be seen as another $\ell_1$ just considering the norm $\|(x^1,\ldots, x^K)\|=\sum_{k=1}^K \|x^k\|_1$. Given a sequence $(x^k)_{k\in \N}\subset \ell_1$ one says that it converges  to $x\in \ell_1$ if $\|x^k-x\|_1\rightarrow 0$. In a direct sum as above,  convergence is simply equivalent to convergence in each of the factors.
We will use the following characterization of convergence for positive sequences in $\ell_1$: given a sequence $(x^k)_{k\in \N}\subset\ell_1$ such that $x^k_n\ge 0$ for all $k,n$,  $\lim_k x^k_n =x_n$ for all $n$, and $\lim_k \sum_{n=-\infty}^\infty x^k_n=\sum_{n=-\infty}^\infty x_n$, then $x^k$ converges in norm to $x\in \ell_1$.

Let us consider $N_f\times N_f$ matrices with values in $\ell_1$. We identify each column of the matrix with $X_1=\oplus_{k=1}^{N_f}\ell_1$, which is another $\ell_1$ . The pointwise determinant $${\rm Det}_1: \underset{N_f}{\underbrace{X_1\times \cdots \times X_1}}\rightarrow \ell_1,$$ is a continuous multilinear map on $X_1$ with values on $\ell_1$. 

We fix $\beta=V\Sigma m$ and consider $\nu$ as the variable (getting sequences in $\nu\in \Z$).

We denote  $$R_L(\nu)=\chi_{[-L,L]}(\nu)\langle \overset{2L+1} {\overbrace{...,\uparrow ,\underset{N_{f}}{\underbrace{\downarrow
,...,\downarrow }}}}%
\rvert \mathrm{e}^{-\beta \hat{H}_{\mathrm{XX}}}T^{-\nu}\lvert \underset{N_{f}}{%
\underbrace{\downarrow ,...,\downarrow }},\uparrow ,...\rangle$$
and 
\begin{equation}\label{eq-ap-1}
R(\nu)=\lim_{L\rightarrow\infty} R_L(\nu)\; (= Z_{\nu}^{\mathrm{eff}}(m))
\end{equation}
From the results in \cite{LS,BC} (see also (\ref{minordet})) we know that both $R_L$ and $R$ are the determinant of a Toeplitz matrix 
\begin{widetext}
$$R_L(\nu)=\chi_{[-L,L]}(\nu)\left|
\begin{matrix}
q^L_\nu(\beta)& q^L_{\nu+1}(\beta) & \cdots & q^L_{\nu+ K-1}(\beta) \\
q^L_{\nu-1}(\beta) & q^L_\nu(\beta) & \cdots & q^L_{\nu+K-2}(\beta)\\
\cdots &\cdots &\cdots &\cdots \\
q^L_{\nu-K+1}(\beta) &q^L_{\nu-K+2}(\beta) &\cdots &q^L_\nu(\beta)
\end{matrix}
\right| \; ,  \quad R(\nu)=\left|
\begin{matrix}
I_\nu(\beta)& I_{\nu+1}(\beta) & \cdots & I_{\nu+ K-1}(\beta) \\
I_{\nu-1}(\beta) & I_\nu(\beta) & \cdots & I_{\nu+K-2}(\beta)\\
\cdots &\cdots &\cdots &\cdots \\
I_{\nu-K+1}(\beta) &I_{\nu-K+2}(\beta) &\cdots &I_\nu(\beta)
\end{matrix}
\right|$$
\end{widetext}
where $I_k(\beta)$ is the Bessel function of imaginary argument and 
$$
q^L_k(\beta) =\frac{1}{2L+1}\sum_{s=-L}^L e^{\frac{2\pi i sk}{2L+1}}e^{\beta\cos \frac{2\pi s}{2L+1}}
$$
which is trivially periodic (in $k$) with period $2L+1$.
As commented in the finite chain analysis in the main text, the first expression is nothing but a Riemann sum associated with the integral representation of the Bessel function $I_k(\beta)$, which shows that $\lim_L q^L_k(\beta) =I_k(\beta)$ for all $k\in \Z$.
Moreover, $I_k(\beta)\ge 0$, $q_k^L(\beta)\ge 0$, and we have the equalities
\begin{equation}\label{eq-ap-2}
\sum_{k=-L}^L q^L_k(\beta)=e^\beta=\sum_{j=-\infty}^\infty I_j(\beta)\; .
\end{equation}
 By the characterization given above,  for any fixed $r\in \Z$, we get that $(\chi_{[-L,L]}(\nu)q^L_{r+\nu}(\beta))_L\subset \ell_1$ converges to $(I_{r+\nu}(\beta))_\nu\in \ell_1$. 
 Now, using the continuity of ${\rm Det}_1$, we get that $R_L$ converges to $R$ on $\ell_1$. This implies trivially weak convergence, that is, for any bounded sequence $(y_\nu)_{\nu\in\Z} $, 
 $$\lim_{L\rightarrow\infty}\sum_{\nu=-\infty}^\infty y_\nu R_L(\nu)= \sum_{\nu=-\infty}^\infty y_\nu R(\nu)\; .$$
 By taking  $y_\nu=e^{i\theta\nu}$ we get
$$Z_{QCD}^{\mathrm{eff}}(\theta)=\sum_{\nu=-\infty}^\infty e^{i\nu \theta} Z_{\nu}^{\mathrm{eff}}(m)=\lim_{L\rightarrow\infty} \sum_{\nu=-L}^L e^{i\nu \theta} R_L(\nu)$$
$$=\lim_{L\rightarrow\infty} \sum_{\nu=-L}^L e^{i\nu \theta}\langle \overset{2L+1} {\overbrace{...,\uparrow ,\underset{N_{f}}{\underbrace{\downarrow
,...,\downarrow }}}}%
\rvert \mathrm{e}^{-\beta \hat{H}_{\mathrm{XX}}}T^{-\nu}\lvert \underset{N_{f}}{%
\underbrace{\downarrow ,...,\downarrow }},\uparrow ,...\rangle
$$
$$= \lim_{L\rightarrow\infty} (2L+1) \langle \overset{2L+1} {\overbrace{...,\uparrow ,\underset{N_{f}}{\underbrace{\downarrow
,...,\downarrow }}}}
\rvert P_\theta \mathrm{e}^{-\beta \hat{H}_{\mathrm{XX}}}P_\theta\lvert \underset{N_{f}}{%
\underbrace{\downarrow ,...,\downarrow }},\uparrow ,...\rangle\;  , $$
which finishes the argument.


\end{document}